\documentclass[preprint, aps,prb]{revtex4-1}

\usepackage{multirow}
\usepackage{braket}
\usepackage{graphicx}
\usepackage{dcolumn}
\usepackage{bm}
\usepackage{array}
\usepackage{color}

\def\bfk{{\bf k}}
\def\bfq{{\bf q}}
\def\bfR{{\bf R}}

\def\bfT{{\bf T}}

\newcommand{\req}[1]{\mbox{Eq.(\ref{#1})}}

\def\smH{\mathcal H}

\def\bfS{{\bf S}}
\def\bfB{{\bf B}}
\def\bfe{{\bf e}}
\def\mub{{\mu_{\rm B}}}
\begin{document}

\title{Nonlinear Extension of the Dynamical Linear Response of Spin: Extended Heisenberg Model}

\author{Haruki Okumura}
\email{h-okumura@issp.u-tokyo.ac.jp}
\affiliation{
Institute of Solid State Physics, The University of Tokyo, 5-1-5 Kashiwanoha, Kashiwa, Chiba, 277-8581, Japan
}

\author{Kazunori Sato}
\affiliation{
  Division of Materials and Manufacturing Science,
  Graduate School of Engineering, Osaka University,
  Osaka, Japan
}
\affiliation{
   Center for spintronics research network (CSRN), Osaka University,
   Osaka, Japan
 }

\author{Takao Kotani}
\email{takaokotani@gmail.com}
\affiliation{
  Advanced Mechanical and Electronic System Research Center, 
  Department of Engineering, Tottori University, Tottori, Japan
}
\affiliation{
   Center for spintronics research network (CSRN), Osaka University,
   Osaka, Japan}

\begin{abstract}
We introduce a new extended Heisenberg model that 
contains orbital-dependent spins 
together with the retarded effects of spin torque. 
The model is directly derived from the dynamical linear
response functions on the transversal spin fluctuation. 
Our model allows us to address effects that are not accessible 
via the usual Heisenberg model.
With the model, we can describe not only the relaxation effects 
due to the Landau damping caused by the Stoner excitations,
but also the nesting effects of the Fermi surface. 
We discuss the possibilities of the extended Heisenberg model
on the basis of high-resolution plots of the spin susceptibility for Fe.
\end{abstract}
\maketitle

\section{Introduction}
We may classify methods of evaluating transversal spin fluctuations 
in first-principles methods into two types.
The first type is based on static linear response methods such as
the Lichtenstein formula \cite{liechtenstein_local_1987,bruno_exchange_2003,vSA99,PhysRevB.64.174402,PhysRevB.103.024418}
or frozen magnon methods.\cite{PhysRevB.55.14975,PhysRevB.58.293,PhysRevB.63.100401}
Under the assumption of the rigid rotation of the spin magnetic moments at atomic sites,
the exchange coupling $J$ is identified to be the second derivative of the total energy
with respect to the rotations.
As a result, we obtain spin waves (SWs) that describes the transversal spin fluctuations.
Although the theory is limited within the linear response in principle, 
we often simply extend it to the nonlinear regime through the Heisenberg model.
Thus, we can determine even magnetic transition temperatures that are not accessible 
within the linear response.

The second type is based on the dynamical linear response theory, 
the so-called ladder approximation in the many-body perturbation theory.
\cite{PhysRevB.21.4118,PhysRevLett.81.2570,PhysRevB.62.3006,di_valentin_spin_2014,andreoni_spin_2018,
skovhus2021dynamic} 
The method can be combined with the quasi-particle self-consistent GW method recently
\cite{okumura_spin-wave_2019,okumura_electronic_2020,kotani08sw,kotani_quasiparticle_2014,kotani_formulation_2015}. 
This type is advantageous in the sense that 
we can take into account the dynamical effects. In addition, we do not need to
assume the rigid rotation. Then, we obtain spin fluctuation spectrum, 
which is not only for SWs but also for the Stoner excitations. 
However, how to extend the dynamical linear response 
to the nonlinear dynamical model has not yet been examined well to the best of our knowledge.

Here we show that we can extend the dynamical linear response to the nonlinear regime through 
an extended Heisenberg model, where we use orbital-dependent retarded functions $J(t-t')$ 
instead of the exchange parameters $J$. 
With the atomistic spin dynamics simulations based on $J(t-t')$,
we can include physical effects that are indispensable in spintronics applications.

\section{Extended Heisenberg model}
Let us recall the Heisenberg model, whose Hamiltonian is given as
\begin{eqnarray}
  \smH = -\frac{1}{2} \sum_{\bfR\ne\bfR'}J_{\bfR\bfR'} \bfS_{\bfR} \cdot \bfS_{\bfR'}+ g\mub \sum_{\bfR} \bfS_{\bfR} \cdot \bfB_{\bfR},
\end{eqnarray}
where we use the Bohr magneton $\mub$ and the g factor $g=2$. $\bfS_{\bfR}$ is the spin operator located at the atomic sites ${\bfR}$.
For simplicity, we neglect the spin-orbit coupling in this paper.
The equation of motion $-i \hbar \dot{\bfS}_\bfR=[\smH,\bfS_\bfR]$ gives
\begin{eqnarray}
\hbar \dot{\bfS}_\bfR(t)= \bfS_\bfR(t) \times \sum_{\bfR'} 
(J_{\bfR\bfR'} \bfS_{\bfR'}(t) - g \delta_{\bfR \bfR'} \mub \bfB_{\bfR'}(t)).
\label{eq:hei0}
\end{eqnarray}
In the case of a general crystal structure, we have $\bfR=(\bfT,a)$, where $\bfT$ is the primitive translation vectors, and
$a$ is the index of sites in the primitive cell.

Here we introduce an extended Heisenberg model as an extension of \req{eq:hei0} as 
\begin{eqnarray}
\hbar \dot{\bfS}_{\bfR m}(t) = \bfS_{\bfR m}(t) \times \left(\sum_{\bfR' m'} \int_{-\infty}^\infty dt'
\left( J_{\bfR m\bfR' m'}(t-t') \bfS_{\bfR' m'}(t') - g \mub \bfB_{\bfR' m'}(t') \delta(t-t')\right)\right),
\nonumber\\
\label{eq:exhei}
\end{eqnarray}
where we use the operator $\bfS_{\bfR m}$ defined as
\begin{eqnarray}
\bfS_{\bfR m}= \sum_{\alpha,\beta} a^\dagger_{j \alpha} \frac{{\boldsymbol{\sigma}}_{\alpha \beta}}{2} a_{i \beta}.
\end{eqnarray}
Here $m=(i,j)$ is a product index for the orbital indices $i$ and $j$ at site $\bfR$; 
the indices $i$ and $j$ denotes the Wannier functions at site $\bfR$.
$\boldsymbol{\sigma}=(\sigma_x,\sigma_y,\sigma_z)$ is the Pauli matrix.
We need retarded functions $J_{\bfR m\bfR' m'}(t-t')$ (=0 for $t-t'<0$) 
for spin dynamics specified by \req{eq:exhei}.
This is the extension of $J_{\bfR\bfR'}$ in \req{eq:hei0}.
We include a term $\bfR=\bfR'$ in the sum of \req{eq:exhei}; constant shifts of diagonal elements $J_{\bfR m\bfR m}(t-t')$ do not affect to the spin dynamics because $\bfS_{\bfR m} (t)\times \bfS_{\bfR m}(t)=0$.
The motion of $\bfS_{\bfR m}(t)$ given by \req{eq:exhei} is the precession 
with the axis of the rotation vector given in the parentheses on the right-hand side of \req{eq:exhei}.
Since $\dot{\bfS}_{\bfR m}(t)$ is perpendicular to ${\bfS}_{\bfR m}(t)$, 
the spin dynamics caused by \req{eq:exhei} is transversal, as well as \req{eq:hei0}.
Because of the retarded response, \req{eq:exhei} specifies a non-Hermitian dynamics showing dissipation.

In this paper, we show that $J_{\bfR m\bfR' m'}(t-t')$ is directly determined by the dynamical linear response theory \cite{okumura_spin-wave_2019}. 
This implies that we can extend the results of the dynamical linear response theory to the nonlinear response
through \req{eq:exhei}. The extended Heisenberg model \req{eq:exhei} contains physical effects that are
not described by the conventional Heisenberg model. 
The orbital-dependent spin moments can rotate independently,
since we treat the orbital degree of freedom by the index $m$.
For example, we can describe a situation in which spins of $e_g$ and of $t_{2g}$ rotate independently.
In addition, the retarded response functions can describe not only the effect of 
friction such as the Gilbert damping, but also the interplay between SWs and the Fermi surface nesting effect.
To illustrate these effects, we have improved our method to calculate the dynamical linear 
responses.
By our method, we can obtain high-resolution dynamical linear responses for Fe.

\subsection{Linear response theory of transversal spin}
Here we summarize the dynamical linear response theory \cite{okumura_spin-wave_2019,andreoni_spin_2018}.
Let us consider ferro-magnetic cases that have finite expectation values
$\langle \bfS_{\bfR m} \rangle$ at the ground state. 
In the linear response theory, we calculate the expectation value $\langle \delta S^+_{\bfR m}(t)\rangle$ 
for a given external field $B^+_{\bfR m}(t)$.
Here we use the fluctuation operator $\delta S^+_{\bfR m}= S^+_{\bfR m} - \langle S^+_{\bfR m} \rangle$;
we assume a collinear case along the z-axis, that is, $\langle S^+_{\bfR m} \rangle=0$, where $S^+=S_x+i S_y$. 

The dynamical linear response is given as 
\begin{eqnarray}
&&\langle \delta S^+_{\bfR m}(t) \rangle= \sum_{\bfR'm'} \int_{-\infty}^\infty 
dt' R^{+-}_{\bfR m \bfR' m'}(t-t') B^+_{\bfR' m'}(t'),
\label{eq:ladder1}\\
&&R^{+-}_{\bfR m \bfR' m'}(t-t')= ((K^{+-})^{-1}-W)^{-1}
\label{eq:ladder2}
\end{eqnarray}
where $K^{+-}_{\bfR m \bfR' m'}(t-t')$ is the kernel matrix that gives the non-interacting 
response without interaction; we see $K^{+-}=R^{+-}$ if $W=0$.
$K^{+-}_{\bfR m \bfR' m'}(t-t')$ is given as the product of up-spin and down-spin Green's functions.
$W$ is the site- and time-diagonal matrix of effective interaction, written as $W_{\bfR m m'} \delta_{\bfR \bfR'}\delta(t-t')$. 

\subsection{Determination of $J_{\bfR m\bfR' m'}(t-t')$}
Here we show that \req{eq:ladder2} is reproduced by the linearization of \req{eq:exhei}.
This is the basis for extending the linear response to the nonlinear regime.

With the linearization by $\bfS_{\bfR m}(t)= \delta \bfS_{\bfR m}(t) + \langle \bfS_{\bfR m} \rangle$ for \req{eq:exhei},
we have 
\begin{eqnarray}
\hbar \delta \dot{\bfS}_{\bfR m}(t) 
&=&\delta \bfS_{\bfR m}(t) \times \left(\sum_{\bfR' m'} \left(\int_{-\infty}^\infty dt J_{\bfR m\bfR' m'}(t)\right) \langle \bfS_{\bfR' m'}\rangle \right) \nonumber \\
&& - \langle \bfS_{\bfR m} \rangle \times g \mub \bfB_{\bfR m}(t) \nonumber \\
&& - \sum_{\bfR' m'} \int_{-\infty}^\infty dt' J_{\bfR m\bfR' m'}(t-t') \langle \bfS_{\bfR m} \rangle \times \delta \bfS_{\bfR' m'}(t').
\label{eq:ehl}
\end{eqnarray}
In the collinear case as $\langle \bfS_{\bfR m} \rangle= S_{\bfR m z}\bfe_z $, \req{eq:ehl} results in
\begin{eqnarray}
\hbar \omega \delta S^+_{\bfR m}(\omega)
&=& \delta S^+_{\bfR m}(\omega) \left(\sum_{\bfR' m'} J_{\bfR m\bfR' m'}(\omega=0) S_{\bfR' m' z}\right) 
\nonumber \\
&& - S_{\bfR m z} g \mub B^+_{\bfR m}(\omega) \nonumber \\
&& - \sum_{\bfR' m'} S_{\bfR m z} J_{\bfR m\bfR' m'}(\omega)  \delta S^+_{\bfR' m'}(\omega).
\label{eq:ehl2}
\end{eqnarray}
\req{eq:ehl2} is rewritten in matrix form as
\begin{eqnarray}
\delta S^+(\omega)
&=& \frac{-\mub g}{{\hbar \omega}S^{-1} - (\sum J(\omega=0)S)S^{-1} + J(\omega) } B^+(\omega),
\label{eq:ehl3}
\end{eqnarray}
where $S$ means the diagonal matrix $S_{\bfR m z}\delta_{\bfR\bfR'}\delta_{mm'}$, whereas\\
$(\sum J(\omega=0)S)=\left(\sum_{\bfR' m'} J_{\bfR m\bfR' m'}(\omega=0) S_{\bfR' m' z}\right) 
\delta_{\bfR\bfR'}\delta_{mm'}$ is diagonal as well.

Let us assume that \req{eq:ehl3}, the linear response based on the extended Heisenberg model \req{eq:exhei}
is in agreement with \req{eq:ladder1}. Then we have
\begin{eqnarray}
{\hbar \omega} {S}^{-1} - (\sum J(\omega=0)S){S}^{-1} + J(\omega) = {-\mub g}((K^{+-})^{-1}-W).
\label{eq:ehlx4}
\end{eqnarray}
\req{eq:ehlx4} is for determining $J_{\bfR m\bfR' m'}(\omega)$ from $(K^{+-})^{-1}-W$.
Since we have no contribution from $\omega$-independent term because of
$\bfS_{\bfR m}(t) \times \bfS_{\bfR m}(t)=0$, as noted at the bottom of the paragraph in \req{eq:exhei},
\req{eq:ehlx4} is reduced to 
\begin{eqnarray}
&&J_{\bfR m\bfR' m'}(\omega) = {-\mub g}((K^{+-})^{-1}-W)_{\bfR m\bfR' m'}(\omega) 
- \frac{\hbar \omega}{S_{\bfR m z}} \delta_{\bfR\bfR'}\delta_{mm'}.
\label{eq:ehoff}
\end{eqnarray}
Generally, the diagonal parts $J_{\bfR m\bfR m}(\omega)\ne 0$, except in the case
that the magnetic moments at $\bfR$ are rotated rigidly as was assumed in the Lichtenstein formula.
Thus we expect $J_{\bfR m\bfR m}(\omega)$ to have a term proportional to $\omega$,
generating a damping torque $\propto \bfS_{\bfR m} \times \dot{\bfS}_{\bfR m}$,
although we should take into account the spin-orbit coupling for accurate calculations
\cite{ke_intersublattice_2019}.
Note that the usual Lichtenstein formula is obtained as an approximation of \req{eq:ehoff} as 
$(K^{+-})^{-1}_{\bfR m \bfR' m'}(\omega=0) \approx \frac{1}{K^{+-}_{\bfR m \bfR m}(\omega=0)}
 K^{+-}_{\bfR m \bfR' m'}(\omega=0) \frac{1}{K^{+-}_{\bfR' m' \bfR' m'}(\omega=0)}$.

\subsection{Physical effects in the extended Heisenberg model}
Our extended Heisenberg model is the natural extension of the conventional Heisenberg model.
We determine $J_{\bfR m\bfR' m'}(\omega)$ of \req{eq:exhei} 
from the dynamical linear responses given by $K^{+-}$ and $W$.
By construction, the linearization of \req{eq:exhei} around the ground state of $\langle S_{\bfR m}\rangle$
exactly reproduces the dynamical linear responses. 
This is essentially the same as in the case of conventional Heisenberg model
where we determine the exchange coupling $J_{\bfR \bfR'}$ using the Lichtenstein formula.

\begin{figure}[ht]
\includegraphics[width=18cm]{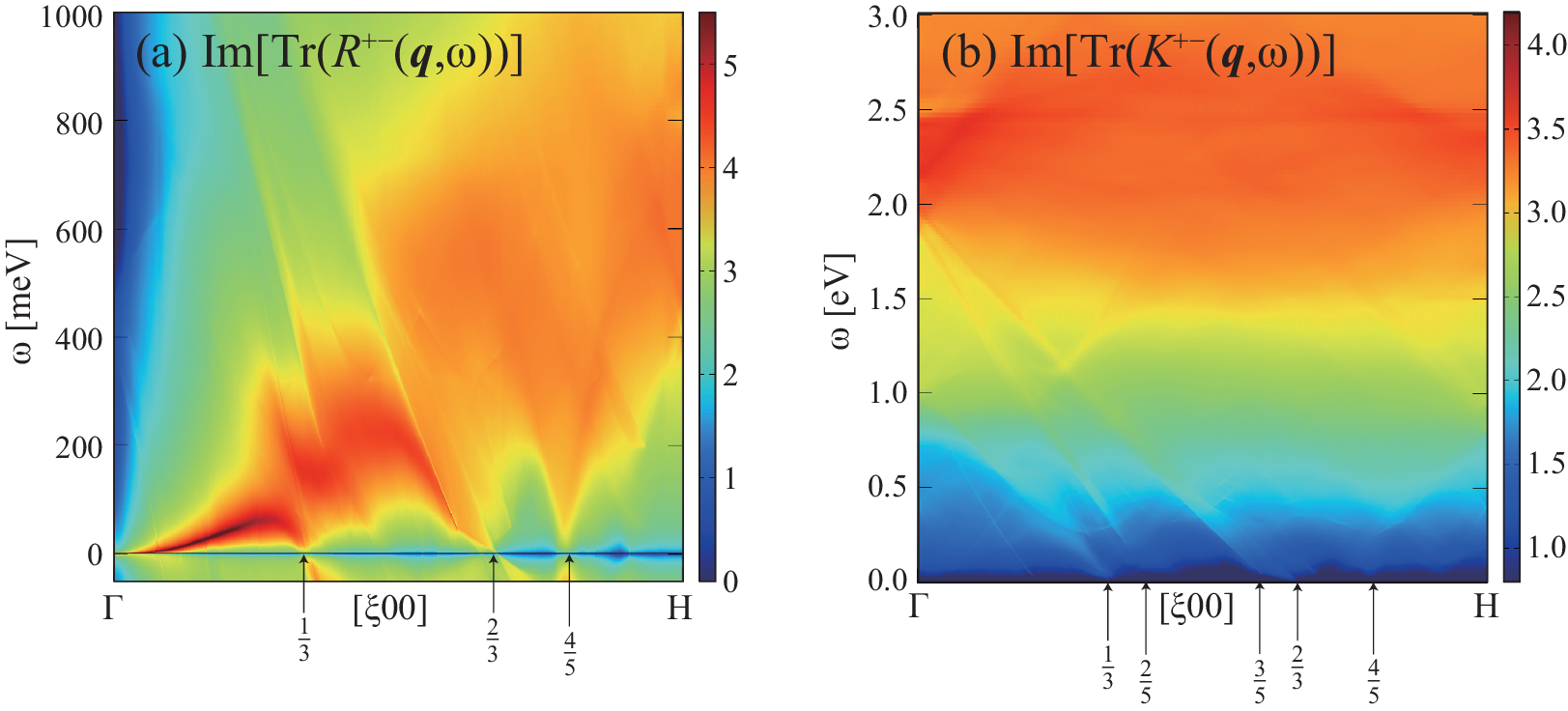}
 \caption{
High-resolution calculation of ${\rm Im}[{\rm Tr}(R^{+-}(\omega))]$, together with 
${\rm Im}[({\rm Tr}(K^{+-}(\omega))]$ for Fe in LDA by the method described 
in Ref.\onlinecite{okumura_spin-wave_2019}. Units in y-axis are in arbitrary logarithmic scales.}
\label{fig:FeLDA}
\end{figure}

Let us examine the physical effects of the extended Heisenberg model 
on the basis of our results of spin susceptibility spectrum ${\rm Im}[{\rm Tr}(R^{+-}(\omega))]$ via \req{eq:ladder2}.
In Fig.\ref{fig:FeLDA}, we show our high-resolution calculation 
of ${\rm Im}[{\rm Tr}(R^{+-}(\omega)]$ for bcc Fe in LDA. 
For the calculation, we have improved a method shown in Ref.\onlinecite{okumura_spin-wave_2019};
we have 100 divisions along $\Gamma$ to H in Fig.\ref{fig:FeLDA}, 
whereas we use $40 \times 40\times40$ divisions in the 1st Brillouin zone for the sum of $\bfk$ in Eq.(12)
of Ref.\onlinecite{okumura_spin-wave_2019}. 
We use the tetrahedron method \cite{rath75} to accumulate the imaginary part of the Stoner excitations.
The tetrahedron method assumes the linear interpolation of 
excitation energies within the microcell of the divisions. We use sufficiently dense  
logarithmic bins for energy for the accumulation. The real part of the Stoner excitation is calculated by
the Hilbert transformation. 
$W$ is corrected a little by a scaling factor $\eta$ 
so as to satisfy the Goldstone theorem \cite{kotani08sw}.
Although there are many such calculations available as shown in the introduction, 
we can clearly see physical effects without confusion, owing to the high resolution of 
Fig.\ref{fig:FeLDA}.

At low energy around $\Gamma$, we see a branch of SWs in Fig.\ref{fig:FeLDA} (a) starting at
$(\bfq=[000],\omega=0)$. The ridge of SWs (= the SW dispersion at $\omega \to 0$) 
is bending towards $(\bfq=[\frac{1}{3} 0 0],\omega=0)$.
This can be interpreted as the effect of the hybridization of the SWs with one of 
the Stoner-excitation boundaries starting at $(\bfq=[\frac{1}{3} 0 0],\omega=0)$. 
We expect commensurate spin-density-wave-like instabilities caused by some perturbations for $\bfq=[\frac{1}{3} 0 0]$. 
We confirmed that we have a self-consistent solution in LDA with the corresponding supercell.
In Fig.\ref{fig:FeLDA} (b), we see corresponding boundary 
starting at $(\bfq=[\frac{1}{3} 0 0],\omega=0)$ and ending at $(\bfq=[000],\omega\sim0.9 {\rm eV})$.

In addition, we see other Stoner excitation boundaries in Fig.\ref{fig:FeLDA} (b).
For example, one of them is starting at $(\bfq=[\frac{3}{5} 0 0],\omega=0)$ and ending at $(\bfq=[000],\omega\sim2 {\rm eV})$.
It seems that $\zeta$ of the several boundaries at $\omega=0$ along the line $[\zeta 0 0]$ 
are simple fractional numbers of integers.
Some of these boundaries are not necessarily linear; for example, we see a parabolic contrast 
of light and deep blue 
at $(\bfq \sim[\frac{3}{10} 0 0],\omega \sim 0.25 {\rm \ eV})$ in Fig.\ref{fig:FeLDA} (b). 
These originate from the superposition of all the
possible spin-flip excitations (= Stoner excitations) for given $(\bfq,\omega)$ calculated 
from the band structures.

In standard textbooks, we expect that the SW dispersion is diminished because of
the Landau damping when the dispersion goes into the Stoner-excitation domain.
However, we rather see strong repulsive hybridizations of the SW dispersion with the Stoner-excitation boundaries.
Corresponding to several boundaries seen in Fig.\ref{fig:FeLDA} (b), we see that the spin fluctuation shown
in Fig.\ref{fig:FeLDA} (a) can be divided into several domains.
Above the domain around $[\frac{2}{3} 0 0]$, we can no longer see clear remnant of the SWs anymore.
Fig.\ref{fig:FeLDA} (a) is the high-resolution version of Fig.7 
in Ref.\onlinecite{di_valentin_spin_2014} by Friedrich et al. 
We see good correspondence between Fig.\ref{fig:FeLDA} (a) and their Fig.7. 
Our Fig.\ref{fig:FeLDA} (a) revealed the fact
that the ``optical branch'' as shown in Fig.8 in Ref.\onlinecite{di_valentin_spin_2014}
should be interpreted by the repulsive hybridization.
Fig.\ref{fig:FeLDA} (a) is consistent with the fact that 
the SWs have not been measured beyond $\sim [\frac{1}{3} 0 0]$ along the $\Gamma-H$ line
by neutron scattering techniques.\cite{lg1985} 

By definition of the extended Heisenberg model, 
we can reproduce Fig.\ref{fig:FeLDA} by the linearization of the extended Heisenberg model.
For $\bfq$ which is a little away from $\Gamma$, we have damping responses that
can be well described by a simple equation of motion of spins such as the Landau-Lifshiz-Gilbert equation.
At larger $\bfq$ (that is, at small $\bfR-\bfR'$), we may need complicated friction mechanism. 
We see some resonances related to the hybridization of the Stone-excitation with SWs. 
The low energy excitations around the starting point of the Stoner-excitation boundaries at $\omega=0$
may have chances to generate some commensurate nonlinear motions of spins.
In the case of Co and so on with two or more atoms in the cell, 
the optical branches of SWs do not necessarily exist 
because of the damping responses as shown in Fig.7 in Ref.\onlinecite{okumura_spin-wave_2019}.
The degree of freedom of orbital-dependent spins may be important for generating frictions 
if the motions of spin magnetic moments are not described well by rigid spin rotations.
The orbital-dependent Heisenberg parameter was analyzed in Ref.\onlinecite{kvashnin_microscopic_2016}.
We expect that the nonlinear motion of spins given by the extended Heisenberg model 
is quite useful in comparison with the conventional Heisenberg model.

To analyze all such nonlinear effects, we may perform
semiquantum dynamics as was performed in Refs.
\onlinecite{barker_semiquantum_2019,Barker_2020}
based on \req{eq:exhei}. We can determine even $T_{\rm c}$ by such a dynamics.
Generally speaking, we can obtain the extended Heisenberg model 
just from any dynamical linear response function.
It is possible to include spin-orbit coupling in the response $R^{+-}$.
Furthermore, we can take into account higher-level electronic correlation 
effects as well as the phonon effect in the evaluation of $R^{+-}$. 
Then we will use the extended Heisenberg model 
as the basis for performing realistic simulations.
Not only the fluctuations of spin magnetic moments 
but also those of the orbital magnetic moments 
might be taken into account along the line of the extended Heisenberg model.


\section{summary}
We have introduced the extended Heisenberg model that is determined by the dynamical linear response.
That is, we have the equation of motion of transverse spins with $J_{\bfR m\bfR' m'}(t-t')$,
which describes the orbital-dependent retarded exchange coupling.

We have discussed what types of physical mechanisms can be included in $J_{\bfR m\bfR' m'}(t-t')$
on the basis of our high-resolution calculation of spin susceptibility for Fe in LDA.
High-resolution calculations clearly reveal physical mechanisms 
that were missing in the low-resolution calculations.
We see not only the damping of SWs, but also the hybridization 
between SWs and the Stoner excitations. 

We believe that we should analyze spin dynamics on the basis of the extended Heisenberg model, 
which plays a key role in connecting first-principles calculations with realistic micro-magnetism simulations 
via atomistic spin dynamics. 
The extended Heisenberg model can be simplified to the Landau-Lifshiz-Gilbert equation.
Such formulation should be useful for the theoretical analysis related to 
spintronics \cite{mahmoud_introduction_2020,hula2021spinwave,Kainuma_2021}.

\begin{acknowledgments}
This work was partly supported by JST CREST Grant number JPMJCR18I2 and by JSPS KAKENHI Grant Number JP18H05212.
T. Kotani acknowledges the support from JSPS KAKENHI Grant Number 17K05499.
\end{acknowledgments}

\bibliography{refkotani}

\end{document}